\documentclass[12pt]{article}
\usepackage{graphicx}

\begin{document}
\begin{center}

\textbf{On the Mass Eigenstate Composition of the\\
$^8$B Neutrinos from the Sun.}

\textbf{A.Kopylov, V.Petukhov}

Institute of Nuclear Research of Russian Academy of Sciences \\
117312 Moscow, Prospect of 60th Anniversary of October Revolution
7A, Russia

beril@al20.inr.troitsk.ru
\end{center}

\begin{abstract}
The present data of gallium experiments provide indirectly the
experimental limit on the fraction of $\nu_2$ mass eigenstate for
the $^8$B neutrinos from the Sun. However, if to use the
experimental data alone, the fraction of $\nu_2$ and, consequently,
$sin^2\theta_{12}$ still is allowed to be varied within a rather
broad range. The further experimental efforts are needed to clear
this point.
\end{abstract}

\textbf{Introduction.}

The study of solar neutrinos performed during more than 40 years in
a number of experiments \cite{1}--\cite{7} have shown unambiguously
that electron neutrinos generated in the matter of the Sun are
converted in other flavors on the way to the Earth. That was the
resolution of a long-standing ``solar neutrino problem''. The
experiment KamLAND \cite{8} has observed the oscillations of
electron neutrinos responsible for the deficit of solar neutrinos,
using antineutrinos from reactors. The global analyses of all data
performed in a number of papers, (see, for example, \cite{9} and
references therein) determined the parameters of neutrino
oscillations responsible for the deficit of solar neutrinos

\begin{equation}
\Delta m^2_{12} = 8.0^{+0.4}_{-0.3}\cdot 10^{-5} eV^2\\
Sin^2\theta_{12} =0.310\pm 0.026
\end{equation}

\noindent at the 68{\%} confidence level. Here symbol 1(2) refers to
a neutrino mass eigenstate with a higher (lower) electron neutrino
component. The neutrino oscillations were observed also in a study
of atmospheric neutrinos \cite{10,11}, here the very specific
signatures of neutrino oscillations in vacuum were demonstrated,
adding to our confidence in a true understanding of this process.
Probably the most interesting manifestation was the observed dip in
the distribution on L/E (length of the pathway to energy of
neutrinos) which can be interpreted only as a neutrino oscillation
in vacuum \cite{10}. Summing up all these data it is impossible to
deny that a great progress has been achieved in these experiments.
But one should be cautious in the final interpretation of the data.
In the uncertainties yielded for the neutrino oscillation parameters
the inherent obscurities of the calculations involved are not always
taken into account. In other words, some regions of the parameter
space are actually excluded not by the experimental data but by the
calculation, which in turn, contains some points still not tested by
experiment. The aim of this paper is to define what can be taken as
a true fact if one is to rely only on the experimental data.

\textbf{Results and Discussions.}

In a recent article \cite{12,13} a thorough analyses has been
performed of the title subject of this paper and it has been shown
that $^8$B solar neutrinos are produced and propagate from the
center of the Sun to the Earth's surface as almost a pure $\nu_2$
mass eigenstate with a purity between 85 and 93{\%}. This result is
in a perfect agreement with all data; hence, it should be beyond of
criticism. But we would like here to draw attention to one point. If
one would have the possibility of measuring in a direct experiment
the mass eigenstate composition of the boron solar neutrinos, this
apparently would put the answer in this question. In fact, there's
no such possibility. The conclusion on the mass eigenstate purity
was formulated on the basis of the measured daytime ratio

\begin{equation}
CC/NC = 0.347\pm 0.038
\end{equation}

\noindent in SNO experiment \cite{7} using the analytical analysis
of the Mikheev-Smirnov-Wolfenstein (MSW) effect \cite{14,15}. The
direct experiment on the neutrino mass eigenstate composition would
have tested that nothing was omitted in the analytical analysis (or
in a numerical one). With the lack of this possibility one cannot
exclude that the composition may turn out to be different from what
is taken now on the basis of this analysis. Here the question
arises: what limitations do come from the experimental data alone?

According to \cite{12} the ratio

\begin{equation}
CC/NC = <P_{ee}> = f_1\cdot cos^2\theta_{12} + f_2\cdot
sin^2\theta_{12}
\end{equation}

\noindent where $f_1$=1 - $f_2$ and $f_2$ are the $\nu_1$ and
$\nu_2$ fractions. One considers here that neutrinos arrive at the
surface of the earth as an incoherent mixture of the neutrino mass
eigenstates \cite{16}. One problem with this model is that the
coherence/incoherence effects have not been observed by experiment,
hence one still can't be confident in a correct consideration of
these effects on a pass of the electron neutrinos from the point of
origin to a very thin resonance layer, where the transition to the
vacuum oscillations occur, and then to the surface of the earth with
all accompanying effects of wave packet spreading and averaging over
space, time and energy. Leaving aside the questions of the
verification of all these moments, we anticipate here a possible
freedom in the evaluation of $f_2$ as it has been done in
\cite{12,13}. This variation of $f_2$ should be certainly concordant
with experiment. If we take CC/NC ratio as a fixed value given by
the SNO experiment and freely vary $f_2$, then we should change
correspondingly $sin^2\theta_{12}$\footnote[1]{A similar procedure
can be performed with the SuperKamioKANDE data, as has been shown in
\cite{13}. Here both experiments can be considered as providing
independent information on $^8$B neutrinos in a full agreement with
each other. For the aims of this paper is not crucial which data to
use; the SNO experiment is somewhat more convenient, because it
utilizes CC and NC channels in the independent measurements on one
and the same installation.}. The question is - what is a limiting
factor in this procedure? The low energy pp neutrinos undergo only
the vacuum oscillations on their flight from the place of origin in
the center of the Sun to the Earth's surface and the studies of
vacuum oscillations by KamLAND and SK \cite{10,11} give us a certain
confidence in a true understanding of this process (although still
some questions can be formulated for the future experiments). The
attenuation coefficient for vacuum oscillations is described by

\begin{equation}
1 - \frac{1}{2}\cdot sin^22\theta _{12}
\end{equation}

This expression does not contain parameters $f_1$, $f_2$ but is a
function of a mixing angle alone. So for each value of $f_2$ one can
get $sin^2\theta_{12}$ and then from the expression (4) get
predictions for a gallium experiment. All other solar neutrino
experiments (the electronic ones) ``see'' only $^8$B neutrinos,
hence the interpretation of their results is dependent on the factor
$f_2$.

\begin{figure}[!ht]
\centering
\includegraphics[width=4in]{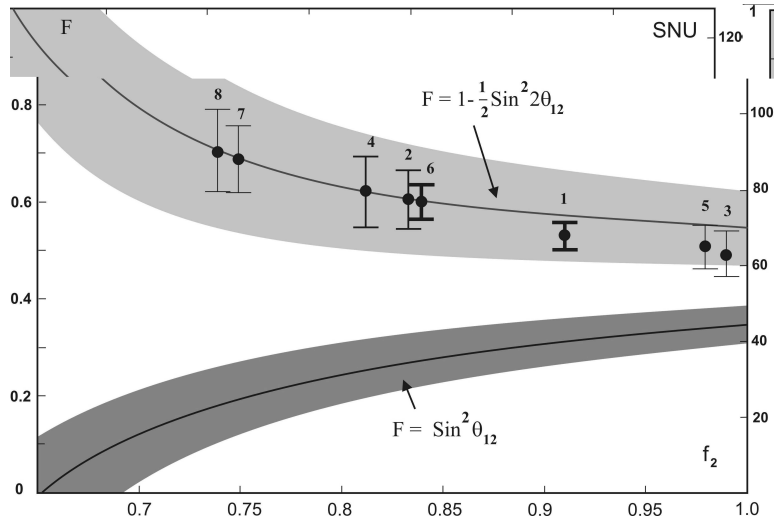}
\caption{The attenuation factor for pp-neutrinos and
$sin^2\theta_{12}$ as a function of $\nu_2$ mass eigenstate
composition: 1 - GALLEX+GNO+SAGE $67.7\pm 3.6$ SNU, 2 - GALLEX
$77.5\pm 6.2$ SNU, 3 - GNO $62.9\pm 5.5$ SNU, 4 - SAGE1 $79.4\pm
8.6$ SNU, 5 - SAGE2 $65.0\pm 5.0$ SNU, 6 - (GALLEX+GNO+SAGE)/$0.88$,
7 - GALLEX/$0.88$, 8 - SAGE1/$0.88$}
\end{figure}

Figure 1 shows the values $sin^2\theta_{12}$ and the resulting
predictions for a gallium experiment as a function of $f_2$ with the
uncertainty determined by the one of CC/NC ratio. The number of
experimental points relates to the central value obtained by
combining the results of SAGE, GALLEX and GNO (1), the high GALLEX
(2) and low GNO (3), the high SAGE1 (4) and low SAGE2 (5) values
obtained within two periods of gallium experiments and the
corresponding points of (1), (2) and (4) divided by 0.88 as a
correction from the combined result of the calibrations from
chromium and argon sources (6-8). The experimental points were taken
from \cite{17}.

The central value (point 1 in Figure 1), as a principal result of
the gallium experiments, is used in a global analyses of all solar
neutrino data plus KamLAND. The allowed $2\sigma$  range for a
mixing angle as determined by Fogli and Lisi in \cite{9} using a
central value is limited by the following expression:

\begin{equation}
0.267 < sin^2\theta_{12} < 0.370
\end{equation}

They investigated also the case when the neutrino potential V(x) is
substituted by $\alpha_{MSW}$V(x), here $\alpha_{MSW}$ is a free
parameter. One can see on Fig.9 of hep-ph/0506083 that for the
central value the preference for standard matter effect
($\alpha_{MSW}$ = 1) is very impressive, but still there is a factor
of 2 uncertainty at $2\sigma$. For defining of the allowed range for
a mixing angle it is very important which points to take from the
results of gallium experiment. If we are to accept the difference
between the results of GALLEX and GNO, and SAGE1 and SAGE2 as the
important one and to take into account the uncertainty with the
calibration of gallium detector, then one should recognize that the
allowed range for $sin^2\theta_{12}$ can be extended to the level of
0.15 and, consequently, the attenuation factor for pp-neutrinos can
approach the value 0.8. To agree with the result of SNO experiment,
the value $f_2$ should be equal to 0.72 in this case. Our point is
that the contradiction of this value and $0.92 \pm 0.02$ calculated
in \cite{12,13} can be explained by some "unknowns" in the
consideration of MSW effect in the matter of the Sun. Whether this
can be the deviation from the standard model ($\alpha_{MSW} \ne 1$)
or not accurate consideration of the neutrino wave packet spreading
with the calculated loss of coherence or something else, this is the
question to the future experiments. One should also take into
account some freedom in the evaluation of the contribution of $^7$Be
neutrinos (and all neutrinos of intermediate energies) to the
neutrino capture rate. At the present time the attenuation factor
for $^7$Be neutrinos is found positioning the $^7$Be peak in the
intermediate range between pure vacuum oscillations and pure matter
conversion. Where exactly the boarder is between these two modes,
this has not been investigated by the experiment so far. So it would
be reasonable to accept as a quite possible outcome, that the
attenuation factor for $^7$Be neutrinos is somewhere between 0.5
(for vacuum oscillations) and 0.35 (for matter conversion). Taking
34.7 SNU as a BSB(GS98) \cite{18} prediction for $^7$Be neutrinos
for gallium we obtain the uncertainty of about 5 SNU.

\begin{center}
\small
\begin{table}[!hb]
\caption{The contribution of different sources to the neutrino
capture rate on chlorine and gallium. As a standard solar model here
it was used BSB(GS98) \cite{18}}
\begin{tabular}
{|p{1.5cm}|p{3cm}|p{3cm}|p{3cm}|} \hline Neutrino \par Source&
Attenuation \par Factor& Cl detector \par SNU BSB(GS98)& Ga detector
\par SNU BSB(GS98)
\\ \hline pp& 0.57& --& 70.2
\\ \hline pep& 0.35 - 0.5& 0.22& 2.8
\\ \hline $^7$Be& 0.35 - 0.5& 1.17& 34.7
\\ \hline $^8$B& 0.35& 6.49& 13.6
\\ \hline $^{13}$N& 0.35 - 0.5& 0.05& 1.9
\\ \hline $^{15}$O& 0.35 - 0.5& 0.16& 2.6
\\ \hline &  Total, SNU& 8.09& 125.8
\\ \hline &  Experiment, SNU& 2.56$\pm$ 0.23& 68.1$\pm$ 3.85
\\ \hline
\end{tabular}
\end{table}
\normalsize
\end{center}

This concerns a chlorine experiment too because, although it is
mainly the detector of $^8$B neutrinos, the $^7$Be neutrinos also
contribute significantly to the total neutrino capture rate in the
chlorine experiment; see Table 1. The current uncertainty in
beryllium neutrinos has a direct consequence that the present result
of the SNO experiment is in a nearly automatic agreement with the
result of a chlorine experiment as one can see from the data of
Table 1. The neutrino capture rate for chlorine detector predicted
by a model BSB(GS98) \cite{18} is 8.1 SNU, from this 6.5 SNU is
given by the boron neutrinos. The rest 1.6 SNU comes from the
neutrinos of intermediate energies. If we are to take the
attenuation factor for boron neutrinos due to the oscillations as
the ratio CC/NC measured by SNO at the central point -- 1$\sigma$,
i.e. 0.31, then we get for the contribution of boron neutrinos in
the chlorine experiment 2.0 SNU. The result of a chlorine experiment
is 2.56 $\pm $ 0.22 \cite{1}. So for all neutrinos of intermediate
energies it is left 0.56 $\pm $ 0.22 SNU, and from here one gets for
the attenuation factor 0.35 $\pm $ 0.14. So even if to take the
central value of a chlorine experiment we have full agreement with
SNO for any mixing angle compatible with SNO, found from the
equation (3). Future BOREXINO \cite{19} and/or KamLAND experiments
can measure the effect from $^7$Be neutrinos, what will be very
helpful in eliminating this uncertainty and, what is also important,
may reveal some contradictions between different solar neutrino
experiments. For gallium experiment the precise measurement of the
effect from beryllium neutrinos will give the possibility to improve
the accuracy in the evaluation of the attenuation factor for
pp-neutrinos, and, consequently, for $sin^2\theta_{12}$, because the
uncertainty of the subtraction of the contribution of beryllium
neutrinos will be substantially reduced.

One can see from Fig.1 that, if it is to come out from the
experimental data alone, i.e. not using a neutrino mass eigenstate
composition suggested by a MSW model, however compelling it is, the
mixing angle still is allowed to be varied over a rather broad
range. At present, gallium experiments provide the only experimental
limit for $sin^2\theta_{12}$ since the attenuation factor for
pp-neutrinos does not depend upon the neutrino mass eigenstate
composition, what is not true for $^8$B neutrinos. Here it is worth
to emphasize that in view of the different results obtained in the
calibration of gallium experiments \cite{17} it appears to be
especially important to get this question settled by a more accurate
calibration measurement. The principal solution of course would be
to measure precisely the mixing angle by the electronic detector of
pp-neutrinos or in a specially dedicated experiment with the reactor
antineutrinos. It has been shown in \cite{20} that the main reason
why Kamland turned out to be rather limited in the determination of
a mixing angle is not an optimal distance from a reactor. If a
detector would be placed in the point of a maximal oscillation
effect (minimum in the oscillation curve) a mixing angle would be
measured with much higher accuracy. It has been shown in \cite{21}
that for the detector (SADO) located in Mt. Komagatake at the
distance 54 km from a reactor complex Kashivazaki-Kariva NPP in
Japan a world-record sensitivity on $sin2\theta_{12} \approx
2{\%}(\approx  3{\%})$ at 68.27{\%} CL can be obtained by 60
GW$\cdot $kt$\cdot $yr (20 GW$\cdot $kt$\cdot $yr) operation. The
important point also is that this experiment (SADO) would provide a
direct prove that the minimum of the oscillation curve really
exists. The present data of KamLAND are very compelling to accept
this as a true fact, but nevertheless it would be nice to
demonstrate this in a direct experiment.

As regards pp neutrinos, the ideal would be to do what SNO
experiment has accomplished using $^8$B neutrinos, i.e. to measure
the flux of electron pp-neutrinos and the total flux of
pp-neutrinos. In spite of a relatively high flux of pp-neutrinos the
task is very difficult because the energy is low ($<$ 0.42 MeV).
This is a great challenge for the present experiment but there are
some ideas and developments aimed at this task. The electron
component of the flux of pp-neutrinos is suggested to be measured
utilizing the charged current interaction of pp-neutrinos with the
nuclei (LENS) \cite{22} and $\nu $ -- e$^{-}$ scattering on Xenon
atoms (XMASS) \cite{23}. Following the strategy realized by SNO
experiment it would be very useful to know also the total flux of pp
neutrinos. It is difficult to suggest an electronic detector which
would measure the total flux of pp-neutrinos via neutral current.
But the problem can be tackled by other way. The total flux can be
found from a luminosity constraint \cite{24}

\begin{equation}
0.913f_{pp}+0.002f_{pep}+0.07f_{Be}+0.015f_{CNO} = 1
\end{equation}

\noindent if we are to find the contribution of non-pp neutrinos
($^7$Be and CNO) to the solar luminosity. The low weights of the
non-pp neutrino-generating thermonuclear reactions in the total
luminosity of the Sun have the consequence that even at a relatively
large uncertainties (10\% for $^7$Be- and 30\% for CNO-neutrinos) in
the measurement of the fluxes of $^7$Be and CNO neutrinos the total
flux of pp-neutrinos can be determined with the precision on the
level of 1\% .Thus, by measuring the flux of CNO neutrinos, a
lithium experiment with 10 tons of lithium \cite{25} can give an
essential ingredient - the total flux of pp-neutrinos generated in
the Sun, which will enable the accuracy in the determination of the
mixing angle to be improved. A non-trivial moment in this
consideration is the following. The lithium detector measures the
flux of electron neutrinos coming to the Earth, i.e. with the
attenuation factor due to the oscillation effect. This factor is the
function of the mixing angle. As the input data, we use the mixing
angle with the presently obtained uncertainties. After evaluation of
the contribution of the CNO cycle to the total luminosity of the
Sun, we can find precisely the total flux of pp-neutrinos. (Here we
assume that, by the time a lithium experiment collects data,
Borexino and KamLAND will measure the flux of $^7$Be neutrinos with
an accuracy of at least 10{\%}, which will enable the contribution
of the $^7$Be neutrino-generated reactions to the total luminosity
of the Sun to be found with an uncertainty $<$ 1{\%}.) Then, by
comparing the flux of pp-neutrinos obtained from the data of the
$\nu $e$^{-}$ scattering experiment (XMASS) with the total flux of
pp-neutrinos, we can find precisely the mixing angle as the result
at the output \cite{25}.

\textbf{Conclusions.}

If to come out from the experimental data alone, the mixing angle
$sin2\theta_{12}$ still is allowed to be varied over a rather broad
interval. This is just the consequence of the fact that by the
present time the suggested analytical analysis of MSW effect (or a
numerical one) has not been tested by experiment and thus can be
considered as a one containing some freedom in the evaluation of the
neutrino mass eigenstates composition. The results of gallium
experiments are of prime importance in limiting a mixing angle,
since the attenuation for pp-neutrinos which undergo the vacuum
oscillations, better studied by experiments, does not depend upon
the neutrino mass eigenstate composition, as a free parameter from
the experimental point of view. The further increase of accuracy in
gallium experiments, particularly a new calibration experiment with
higher accuracy, can improve the current situation substantially.
The higher is the neutrino capture rate in gallium the lower is
$f_2$ and $sin^2\theta _{12}$ as one can see from Fig.1. The further
progress can be achieved by means of the electronic detector of
pp-neutrinos or by a dedicated experiment with the reactor
antineutrinos. The measurement of the fluxes of CNO neutrinos can be
helpful to determine precisely a total flux of pp-neutrinos through
a luminosity constraint. A lithium radiochemical detector on 10 tons
of lithium appears to hold promise for the solution of this task.

No attempts were made here to determine the possible 2$\sigma $,
3$\sigma $ regions for $sin^2\theta _{12}$ or correspondingly, for
$f_2$. Our conclusion is that we are still in the phase when it is
premature to define precisely these parameters. New experiments are
needed to clarify the points. We share the belief with other
physicists, that everything is fine with MSW mechanism, as it has
been used in the analysis of the experimental data, and that future
experiments will prove this. However, the substantial difference of
our position is that we insist on the necessity to test this
mechanism in a direct experiment prior to define with a high
precision the parameters, which so far belong to a rather broad
field of uninvestigated possibilities.

\textbf{Acknowledgements.}

This work was supported in part by a Russian Foundation for Basic
Research (project no. 04-02-16678), by the grant of Russia Leading
Scientific Schools N 5573.2006.02 and by the Program of fundamental
research of Presidium of Russian Academy of Sciences ''Neutrino
Physics''.

\textbf{Appendix.}

Here we would like to give the answers to the typical questions
asked during the discussions since the first publication of our
preprint in \cite{26}.

\textbf{Question}: Many groups throughout the world made the
combined analyses of the solar and KamLAND data and yielded the
parameters of neutrino oscillations with high accuracy in a full
agreement with each other. How to understand it, were all they wrong
while you are right?

\textbf{Answer:} We don't say that they are wrong. All these
publications are certainly very illuminating, and the published
results have the guidance for the future experiments, but they were
obtained in the supposition that the calculation by MSW model is
absolutely correct, which is not obvious. We think that more
attention should be paid to the fact that, when we apply to MSW
mechanism, we get a new parameter -- $f_2$, and one can gain large
uncertainties in the evaluation of the neutrino mass eigenstate
composition due to the internal obscurities associated with the
application of MSW mechanism which still has not been tested by
experiment. As the consequence we make a statement that allowed
region for $sin^2\theta _{12}$ is in fact broader than is expected
from the results of the global analyses of data and that future
experiments may bring us a surprise.

\textbf{Question}: Some experiments in your paper are definitely
belittled (chlorine, SK) while others raised in significance (SNO
and gallium). In the global analyses of data they are of comparable
importance. Large uncertainties in the evaluation of $f_2$ and of a
mixing angle may have a simple explanation: you just don't use all
data, i.e. your estimates are too schematic.

\textbf{Answer:} We operate mainly with the data of gallium and SNO
experiments because they give information both for low (where we
have the vacuum oscillations) and high (where we have the neutrino
conversion) energies, i.e. for pp- and Boron-neutrinos. We need the
information for low and high energy, because we see the possibility
to extend the allowed region for $sin^2\theta _{12}$ resolving the
system of two equations with two unknown parameters: $f_2$ and
$sin^2\theta _{12}$. Other experiments (chlorine and SK) add
information for Boron-neutrinos, but give no information for
pp-neutrinos. The inclusion of chlorine and SK data diminish a
little the uncertainty, of course, but being the experiments which
just give the independent information on boron neutrinos (as
explained also in the text of the manuscript) they don't change the
situation drastically. The main effect is achieved here not due to
some resulting loss of the accuracy of the global analyses, but
because the essential ingredient gets lost -- the limiting power of
the calculation, coming from the resonant character of the MSW
effect. In other words, a certain regions ($f_2)^{\prime}$ (outside
of the claimed interval 91 $\pm $ 2 {\%}) and, consequently, of a
mixing angle (sin$^2\theta _{12})^{\prime}$, obtained from the
expression (3), are excluded not by the experimental data, but
because in the calculation of MSW one will not get these ``extreme''
regions ($f_2)^{\prime}$, as one can see from the comparison of
Fig.1 of our paper and Fig.1 in \cite{12}.

It is impossible to underestimate the role of chlorine and SK
experiments in the study of solar neutrinos. The chlorine experiment
was first to discover the deficit of boron neutrinos, and this
number -- 3, obtained for the attenuation of boron neutrinos, was
later confirmed by Kamiokande, SK an SNO. SuperKamiokande made the
most precise measurement of elastic scattering for Boron neutrinos,
with all spectral information, what is very important. Moreover, SK
studied the neutrino oscillations for atmospheric neutrinos, and the
dip in L/E distribution observed by SK is by the present time
probably the most powerful manifestation of the neutrino
oscillations in vacuum.

\textbf{Question}: Well, probably one could agree with your
statement, that so far MSW effect has not been tested by experiment,
we can not take the results with absolute reliability. But this are
not big news for the audience. Of course, people understand it. But
obviously, what is more important now, the global analyses of all
data performed by a number of groups, using the MSW mechanism
developed by authors, which gives good results, wholly
self-consistent. People believe that tomorrow a new experiment will
measure the mixing angle for vacuum oscillations with a good
accuracy and it will be in agreement with the present one, found
from MSW. This will prove that MSW is correct and this will be the
end of the story.

\textbf{Answer:} If tomorrow a new experiment will prove that MSW is
correct in all details, this will be just excellent. Nature is
following to the prescriptions of the theoreticians. What can be
better than this? But if we are physicists, we should clearly
separate what we believe in from what we know. The aim of this paper
is exactly in this. Certainly, it is a beautiful thing: it looks
like a magic that electron neutrinos born in the Sun come to the
Earth as almost pure (91 $\pm $ 2 {\%}) $\nu_2$ mass eigenstate. But
one should agree that this is not the experimental fact. This is the
result of the calculation which, in turn, contains some obscurities,
still not tested by experiment. So why should we believe that this
number is correct? Moreover, this result does not stand by itself;
it is incorporated in the calculation of the neutrino oscillation
parameters. The uncertainties in the determination of $f_2$ (if
there are any) produce additional uncertainties for
$sin^2\theta_{12}$.

\end{document}